\documentclass[a4paper]{jpconf}

\usepackage{amssymb}
\usepackage{amsmath}
\usepackage{afterpage}
\usepackage{graphicx}
\usepackage{color}

\newcommand{\bea}{\begin{eqnarray}}
\newcommand{\eea}{\end{eqnarray}}
\newcommand{\be}{\begin{equation}}
\newcommand{\ee}{\end{equation}}
\renewcommand{\tr}{\mathop{\mathrm{Tr}}}

\newcommand{\p}{{\bf p}}

\newcommand{\0}{{\bf 0}}

\newcommand{\nom}{{\nonumber}}

\newcommand{\expm}{e^{-\beta\left(E_p-\mu_q\right)}}

\newcommand{\expmmm}{e^{-3\beta\left(E_p-\mu_q\right)}}
\newcommand{\expp}{e^{-\beta\left(E_p+\mu_q\right)}}
\newcommand{\expppp}{e^{-3\beta\left(E_p+\mu_q\right)}}

\renewcommand{\eref}[1]{Eq.~\eqref{#1}}
\newcommand{\secref}[1]{Sec.~\ref{#1}}

\allowdisplaybreaks

\begin{document}

\title{Chiral phase transition in the vector meson extended linear
  sigma model } \author{P{\'e}ter Kov\'acs$^1$, Zsolt Sz{\'e}p$^2$ and
  Gy{\"o}rgy Wolf$^1$} \address{$^1$ Institute for Particle and
  Nuclear Physics, Wigner Research Centre for Physics, Hungarian
  Academy of Sciences, H-1525 Budapest, Hungary} \address{$^2$
  MTA-ELTE Statistical and Biological Physics Research Group, H-1117
  Budapest, Hungary}

\ead{kovacs.peter@wigner.mta.hu}

\begin{abstract}
  In the framework of an SU(3) (axial)vector meson extended linear
  sigma model with additional constituent quarks and Polyakov loops,
  we investigate the effects of (axial)vector mesons on the chiral
  phase transition. The parameters of the Lagrangian are set at zero
  temperature and we use a hybrid approach where in the effective
  potential the constituent quarks are treated at one-loop level and
  all the mesons at tree-level. We have four order parameters, two
  scalar condensates and two Polyakov loop variables and their
  temperature and baryochemical potential dependence are determined
  from the corresponding field equations. We also investigate the
  changes of the tree-level scalar meson masses in the hot and dense
  medium.
\end{abstract}

\section{Introduction}

Forthcoming heavy ion experiments, such as the planned CBM experiment
at FAIR will explore the QCD phase diagram in the high density and
moderately high temperature region. This region is very interesting
since it is believed that the critical endpoint (CEP) -- which
separates the crossover and first order phase transition regions along
the phase boundary -- if it exists, should be found there.

Earlier experiments such as RHIC, SPS, or LHC investigated the low
density, high temperature region. No sign of the chiral phase
transition has been observed experimentally, which is not so
surprising considering that the phase transition is of crossover type
there, which is very hard to observe experimentally. Theoretically, if
the transition is of first order -- and the system is infinite --
certain quantities have discontinuities crossing the phase boundary,
which gives a much higher chance to observe it. Although, since {in
  real physical experiments} the system is always finite, there will
not be any observable discontinuities, but just peaks.

Due to the difficulties of the observation it is very important to
gather as much theoretical information as possible from the phase
boundary, in order to assist future experiments. According to that, we
would like to investigate the chiral phase transition in the framework
of a (axial)vector meson extended linear sigma model with additional
constituent quarks and Polyakov loops.

The paper is organized as follows. The model is briefly
presented in \secref{Sec:model} and its parametrization is discussed
in \secref{Sec:param}. The four field equations which provide
the temperature and chemical potential dependence of the order
parameters are given in \secref{Sec:field_eqns}. Finally, 
the results are discussed in \secref{Sec:result}, where we also conclude.

\section{The Model}
\label{Sec:model}
In terms of matrix valued (pseudo)scalar and (axial)vector meson
fields the Lagrangian is,
\begin{align}
  \mathcal{L} & = \tr[(D_{\mu}\Phi)^{\dagger}(D_{\mu}\Phi)]-m_{0}%
  ^{2}\tr(\Phi^{\dagger}\Phi)-\lambda_{1}[\tr(\Phi^{\dagger}%
  \Phi)]^{2}-\lambda_{2}\tr(\Phi^{\dagger}\Phi)^{2}\nom \\
  & -\frac{1}{4}\tr(L_{\mu\nu}^{2}+R_{\mu\nu}^{2})+\tr\left[ \left(
      \frac{m_{1}^{2}}{2}+\Delta\right)
    (L_{\mu}^{2}+R_{\mu}^{2})\right]
  +\tr[H(\Phi+\Phi^{\dagger})]\nom \\
  & +c_{1}(\det\Phi+\det\Phi^{\dagger})+i\frac{g_{2}}{2}(\tr
  \{L_{\mu\nu}[L^{\mu},L^{\nu}]\}+\tr\{R_{\mu\nu}[R^{\mu},R^{\nu
  }]\})\nom \\
  & +\frac{h_{1}}{2}\tr(\Phi^{\dagger}\Phi)\tr(L_{\mu}
  ^{2}+R_{\mu}^{2})+h_{2}\tr[(L_{\mu}\Phi)^{2}+(\Phi R_{\mu}
  )^{2}]+2h_{3}\tr(L_{\mu}\Phi R^{\mu}\Phi^{\dagger})\label{Eq:Lagr}\\
  & +g_{3}[\tr(L_{\mu}L_{\nu}L^{\mu}L^{\nu})+\tr(R_{\mu}R_{\nu
  }R^{\mu}R^{\nu})] + g_{4}[\tr\left( L_{\mu}L^{\mu}L_{\nu}L^{\nu
    }\right) \nom\\
  & + \tr\left( R_{\mu}R^{\mu}R_{\nu}R^{\nu}\right)] + g_{5}\tr\left(
    L_{\mu}L^{\mu}\right) \,\tr\left(
    R_{\nu}R^{\nu}\right) + g_{6} [\tr(L_{\mu}L^{\mu})\,\tr(L_{\nu}L^{\nu})\nom \\
  & + \tr(R_{\mu}R^{\mu})\,\tr(R_{\nu}R^{\nu})] + \bar{\Psi}i
  \gamma_{\mu}D^{\mu}\Psi - g_{F}\bar{\Psi}\left(\Phi_{S} +
    i\gamma_5\Phi_{PS}\right)\Psi,\nom % \\
\end{align}
where $D^{\mu}\Phi = \partial^{\mu}\Phi-ig_{1}(L^{\mu}\Phi-\Phi
R^{\mu})-ieA_{e}^{\mu}[T_{3},\Phi],$ $L^{\mu\nu}
= \partial^{\mu}L^{\nu}-ieA_{e}^{\mu}[T_{3},L^{\nu}] -
\{\partial^{\nu}L^{\mu} - ieA_{e}^{\nu}[T_{3},L^{\mu}]\},$
$R^{\mu\nu} = \partial^{\mu}R^{\nu} - ieA_{e}^{\mu}[T_{3}, R^{\nu}]
- \{ \partial^{\nu}R^{\mu} -
  ieA_{e}^{\nu}[T_{3},R^{\mu}]\},$ and $D^{\mu} = \partial^{\mu}
- iG^{\mu}.$ The field content of the Lagrangian is as 
follows, $\Phi$ is the scalar/pseudoscalar field, $L^{\mu}$ and
$R^{\mu}$ are the left and right handed vector fields, 
$\Psi=(u,d, s)^{\text{T}}$ stands for the constituent quark fields,
$G^{\mu}$ is the gluon field, while $H$ is the external
field. As usual, the nonstrange and strange scalar fields are 
shifted by their expectation values $\phi_{N}$ and $\phi_S$
(scalar condensates). A previous version of this model, without the
constituent quarks and Polyakov loops and with a different anomaly
term was soundly analyzed at zero temperature in \cite{elsm_2013}.

\section{Determination of the parameters of the Lagrangian }
\label{Sec:param}

In \eref{Eq:Lagr}, not considering the unused $g_3, g_4, g_5,$ and
$g_6$, there are $14$ unknown parameters, namely $m_0$ -- the bare
(pseudo)scalar mass, $\lambda_1$, and $\lambda_2$ -- the
(pseudo)scalar self-couplings, $c_1$ -- the $U_{A}(1)$ anomaly
coupling, $m_1$ -- the bare (axial)vector mass, $h_1$, $h_2$ and $h_3$
-- the (axial)vector--(pseudo)scalar couplings, $\delta_S$ -- the
(axial)vector explicit break coupling, $\phi_N$ and $\phi_S$ -- the
scalar condensates, $g_F$ -- the Yukawa coupling, and $g_1$ and $g_2$
-- two (axial)vector couplings. Compared to the parametrization done
in \cite{elsm_2013}, in the present case we have two modifications: i)
there are two additional equations for the constituent quark masses,
$m_{u/d} = g_F\phi_N/2$ and $m_s = g_F\phi_s/\sqrt{2}$, for which we
use the values $m_{u/d} = 330$~MeV and $m_{s} = 500$~MeV; ii) the
inclusion of the fermion vacuum fluctuations modifies at $T=\mu=0$ the
masses and decay widths of the model from which the parameters were
calculated in \cite{elsm_2013}. This is because in the present
approach the masses used in the parametrization are the curvature
masses which are the second derivatives of the grand canonical
potential with respect to the meson fields at the minimum. The
curvature masses consist of two parts: a usual tree-level part, which
can be found in \cite{elsm_2013}, and the vacuum and thermal
contributions of the one-loop fermion potential (bosonic fluctuations
are neglected). The later one can be found in \cite{tiwari_2013,
  chatterjee_2012} with Polyakov loops, and in \cite{schaefer_2009}
without Polyakov loops. Scanning through the parameter space,
curvature masses and decay widths are calculated and compared using a
$\chi^2$ minimization method \cite{MINUIT} to the experimental data
taken from the PDG \cite{PDG} and the above values of $m_{u/d}$ and
$m_{s}$, as described in \cite{elsm_2013}.

It is important to note, that in the scalar sector there are more
physical particles than we can describe with one $q\bar q$ nonet, as
in nature there are two $a_0$, two $K_0^{\star}$ and five $f_0$
particles (see \cite{PDG}). Since in one scalar nonet there are one
$a_0$, one $K_0^{\star}$ and two $f_0$'s, we have $2\cdot 2\cdot
\binom{5}{2} = 40$ different possibilities for matching the scalar
sector with physical particles. However, we consider here only cases
where $a_0$ and $K_0^{\star}$ correspond to the $a_0(980)$ and
$K_0^{\star}(800)$ physical particles.

We apply two different parametrization scenarios here. In the first
one we do not fit the very uncertain isoscalars ($f_0^L$, $f_0^H$),
consequently $m_0$ and $\lambda_1$ always appear in the same
combination $C_{1} = m_{0}^{2} + \lambda_{1} \left(\phi_{N}^{2} +
  \phi_S^2 \right)$ in all the expressions, thus we can not determine
them separately. A similar combination $C_{2} = m_{1}^{2}+
\frac{h_{1}}{2} \left(\phi_{N}^{2} + \phi_{S}^{2}\right)$ appears in
the vector sector for $m_1$ and $h_1$ \cite{elsm_2013}.  Practically,
we choose $\lambda_1=h_1=0$ in this scenario. Even if we do not fit
the isoscalars, we still have the possibility to get parametrizations
with different values of $m_{f_0^L}$, which has a huge effect, as will
be seen immediately, on the finite $T/\mu_B$ behavior. We consider two
subcases, labeled by '$1a$' and '$1b$': one with a high and one with a
low value of $m_{f_0^L}$, that is 1326~MeV and 402~MeV, respectively.
In the second scenario, labeled as case '$2$', we fit the isoscalars to
$f_0(500)$ and $f_0(1370).$ The parameter values for case '$1b$' and
case '$2$' are given in Table~\ref{tab-param}.
\begin{table}[!t]
  \centering
  \caption{Parameters determined by $\chi^2$ minimization in the two cases}
  \lineup
  \label{tab-param}   
  \begin{tabular}{lll||lll}
    \hline
    Parameter & Value($1b$) & Value($2$) &
    Parameter & Value($1b$) & Value($2$) \\\hline
    $\phi_{N}$ [GeV] & \0$0.1359 $ & \0$0.1333 $ & $h_{2}$ & $4.8765 $ & $2.7065 $\\
    $\phi_{S}$ [GeV]& \0$0.1400 $ & \0$0.13823 $ & $h_{3}$ & $4.6523 $ & $3.7935 $\\
    $m_{0}^2$ [GeV$^2$] & \0$\-0.0103 $ & \0$0.0394 $ & $\delta_{S}$ [GeV$^2$] & $0.1114 $ & $0.1178 $ \\
    $m_{1}^2$ [GeV$^2$] & \0$0.5600 $ & \0$0.5508 $ & $c_{1}$ [GeV] & $1.5293 $ & $1.600 $ \\
    $\lambda_{1}$ & \0$0$ (undetermined) & \0$\-1.2200$ & $g_{1}$ & $5.5737 $ & $5.5761 $\\
    $\lambda_{2}$ & $23.0638 $ & $23.7957 $ & $g_{2}$ & $2.1263 $ & $1.3889 $\\
    $h_{1}$ & \0$0$ (undetermined) & \0$2.6007$ & $g_{F}$ & $4.3650 $ & $4.4217 $ \\\hline
  \end{tabular}
\end{table}
% ----------------------------------------- file:
% param_gyuri_c1_low_sigma.dat (1st)
% ----------------------------------------- # phiN 0.13585346032037868
% # phiS 0.14002839985993989 # m0^2 -1.0289873450119537E-002 # m1^2
% 0.55999985642577654 # L1 0.0000000000000000 # L2 23.063764519511889
% # xhi1 0.0000000000000000 # xhi2 4.8764947686978344 # xhi3
% 4.6523057282678479 # des 0.11137124085996086 # cc1
% 1.5292832423435896 # g1 5.5737089455082183 # g2 2.1262598410700302 #
% gfe 4.3650041704192972

% --------------------------------------------------------- file:
% param_a0_0.9800_Ks_0.6760_f0L_0.4750_f0H_1.3500.dat
% --------------------------------------------------------- # phiN
% 0.13330539628608781 # phiS 0.13822796155382702 # m0^2
% 3.9418084731501532E-002 # m1^2 0.55081315420681121 # L1
% -1.2200418537356958 # L2 23.795670355766376 # xhi1
% 2.6006510357548609 # xhi2 2.7065427535503050 # xhi3
% 3.7935211587084479 # den 0.0000000000000000 # des
% 0.11780784247429077 # cc1 1.6004622724211652 # cc2
% 0.0000000000000000 # ccm 0.0000000000000000 # g1 5.5761056771081128
% # g2 1.3889215751903594 # gfe 4.4217108421320290

\section{$T/\mu_B$ dependence of the order parameters 
and curvature masses }
\label{Sec:field_eqns}

In medium, the values of the order parameters -- which are the two
scalar condensates $\phi_N$ and $\phi_S$ and the two Polyakov loop
variables $\Phi$ and $\bar{\Phi}$ -- change with the
temperature/chemical potential. The two scalar condensates encodes the
effect of both spontaneous and explicit symmetry breakings, while the
Polyakov loop variables mimic some properties of the quark
confinement, which naturally emerge in mean field approximation, if
one calculates free fermion grand canonical potential on a constant
gluon background (for more details see \cite{Fukushima:2003fw}).

In order to determine the $T/\mu_B$ dependence of the order parameters
and curvature masses, we use four coupled stationarity equations
(field equations), which require the vanishing of the first
derivatives of the grand canonical potential with respect to the order
parameters.  We apply here a hybrid approach, where we only consider
vacuum and thermal fluctuations for the fermions and not for the
bosons. In this case the equations are given by
\begin{align}
  -\frac{d }{d \Phi}\left( \frac{U(\Phi,\bar\Phi)}{T^4}\right) +
  \frac{2 N_c}{T^3}\sum_{q=u,d,s} \int \frac{d^3 \p}{(2\pi)^3}
  \left(\frac{e^{-\beta E_q^{-}(p)}}{g_q^-(p)} + \frac{e^{-2\beta
        E_q^{+}(p)}}{g_q^+(p)}
  \right) &= 0,\label{eq_Phi}\\
  -\frac{d}{d \bar\Phi}\left( \frac{U(\Phi,\bar\Phi)}{T^4}\right) +
  \frac{2 N_c}{T^3}\sum_{q=u,d,s} \int \frac{d^3 \p}{(2\pi)^3}
  \left(\frac{e^{-\beta E_q^{+}(p)}}{g_q^+(p)} + \frac{e^{-2\beta
        E_q^{-}(p)}}{g_q^-(p)}
  \right) &= 0,\label{eq_Phibar}\\
  m_0^2 \phi_N + \left(\lambda_1 + \frac{1}{2} \lambda_2 \right)
  \phi_N^3 + \lambda_1 \phi_N \phi_S^2 - h_N
  +\frac{g_F}{2}N_c\left(\langle u{\bar u}\rangle_{_{T}} + \langle
    d{\bar
      d}\rangle_{_{T}} \right) &= 0,\label{eq_phiN}\\
  m_0^2 \phi_S + \left(\lambda_1 + \lambda_2 \right) \phi_S^3 +
  \lambda_1 \phi_N^2 \phi_S - h_S +\frac{g_F}{\sqrt{2}}N_c \langle
  s{\bar s}\rangle_{_{T}} &= 0,\label{eq_phiS}
\end{align}
where $U(\Phi,\bar\Phi)$ is the Polyakov loop potential, for which we
used a polynomial form with coefficient taken from \cite{ratti_2006},
and
\begin{align*}
  g_q^+(p) &= 1 + 3\left( \bar\Phi + \Phi e^{-\beta E_q^{+}(p)}
  \right)
  e^{-\beta E_q^{+}(p)} + e^{-3\beta E_q^{+}(p)}, \\
  g_q^-(p) &= 1 + 3\left( \Phi + \bar\Phi e^{-\beta E_q^{-}(p)}
  \right)
  e^{-\beta E_q^{-}(p)} + e^{-3\beta E_q^{-}(p)}, \\
  E_q^{\pm}(p) &= E_q(p) \mp \mu_B/3,\; E_{u/d}(p) = \sqrt{p^2 +
    m_{u/d}^2},\; E_{s}(p) = \sqrt{p^2 + m_{s}^2}, \\
  \langle q{\bar q}\rangle_{_{T}} &= -4m_q \int \frac{d^3
    \p}{(2\pi)^3}\frac{1}{2E_q(p)}\left(1 - f^-_\Phi(E_q(p)) -
    f^+_\Phi(E_q(p))\right),
\end{align*}
with the modified Fermi\,--\,Dirac distribution functions
\begin{align}
  f^+_\Phi(E_p) & =\frac{ \left( \bar\Phi + 2\Phi \expm \right) \expm
    + \expmmm } {1 + 3\left( \bar\Phi + \Phi \expm \right) \expm +
    \expmmm}, \nom\\
  f^-_\Phi(E_p) & =\frac{ \left( \Phi + 2 \bar\Phi \expp \right) \expp
    + \expppp }{1 + 3\left( \Phi + \bar\Phi \expp \right) \expp +
    \expppp}. \nom
\end{align}

\section{Results and Conclusion}
\label{Sec:result}

\begin{figure}[!t]
  \centering
  \begin{minipage}{.48\textwidth}
    \centering
    \includegraphics[width=1.1\textwidth]{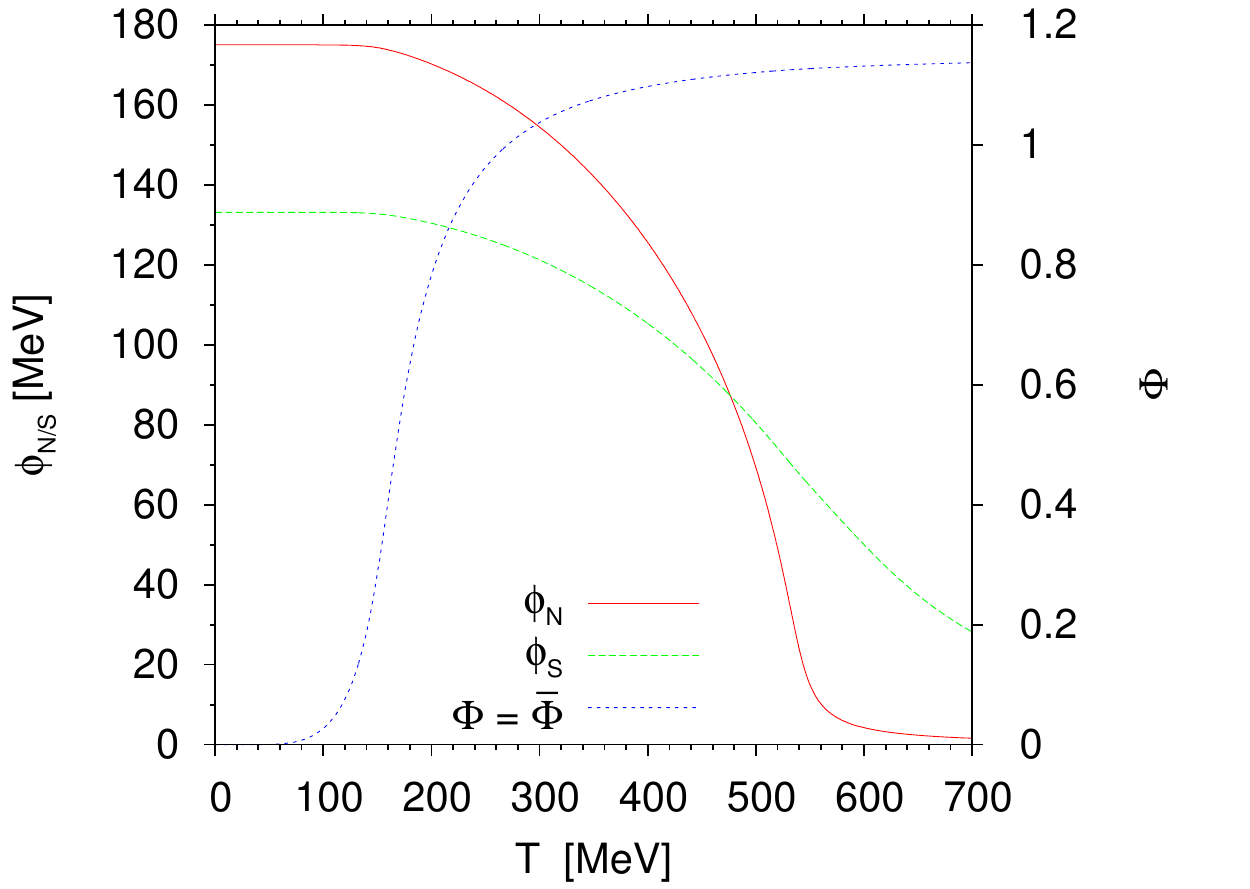}
    \caption{Temperature dependence of the order parameters with the
      parametrization '$1a$' ($m_{f_0^L}=1326$~MeV).}
    \label{order_hi}
  \end{minipage}
  \hspace*{0.02\textwidth}
  \begin{minipage}{.48\textwidth}
    \centering
    \includegraphics[width=1.1\textwidth]{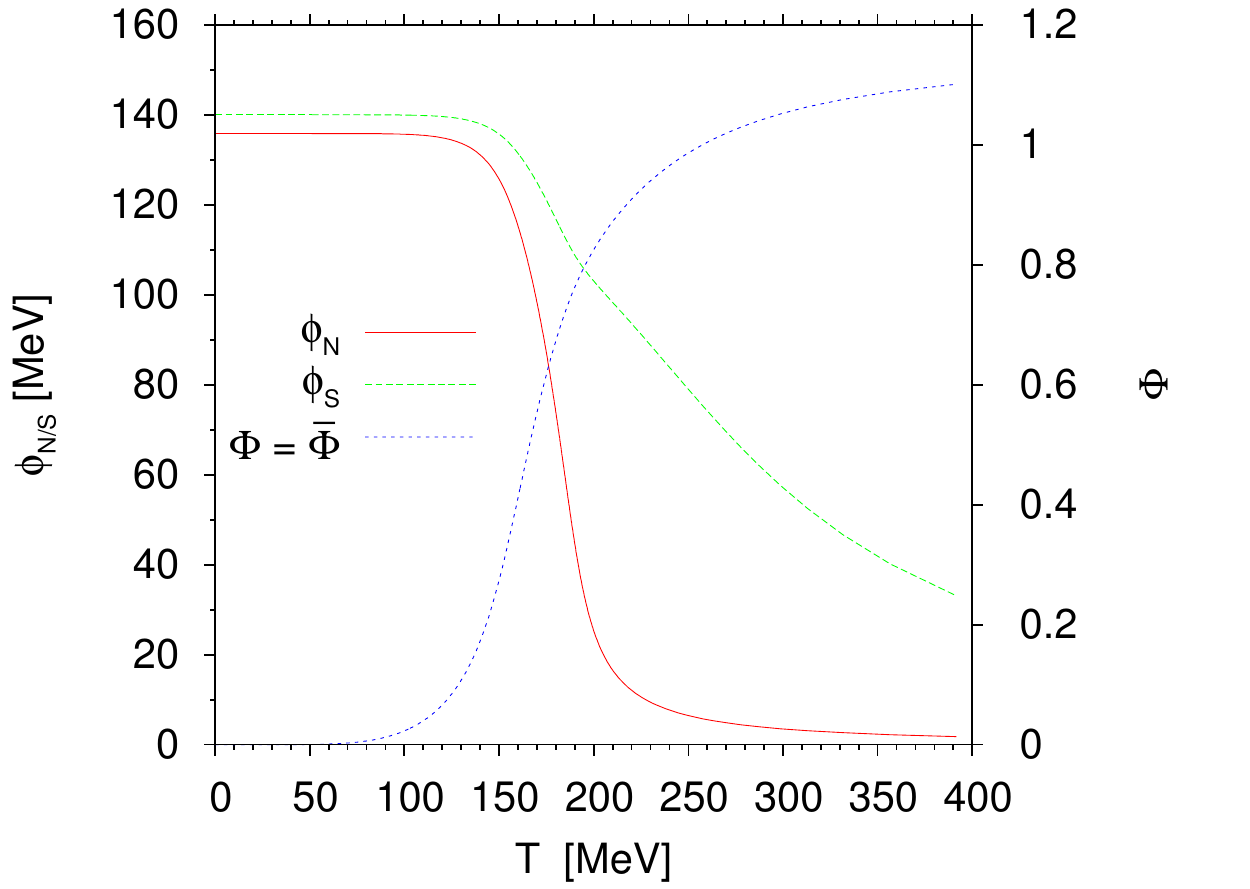}
    \caption{Temperature dependence of the order parameters with the
     parametrization '$1b$' ($m_{f_0^L}=402$~MeV).}
    \label{order_lo}
  \end{minipage}
\end{figure}

By solving the system of Eqs.~\eqref{eq_Phi}-\eqref{eq_phiS}, the
temperature and baryochemical potential dependence of the order
parameters and the curvature masses can be determined. In
Fig.~\ref{order_hi} the order parameters can be seen with the
parametrization '$1a$' at $\mu_B=0$. Here the pseudocritical temperature
($T_c$) is around $550$~MeV, which is much higher than the continuum
lattice result \cite{aoki_2006}, which is $T_c=151$~MeV.
In Fig.~\ref{order_lo} the order parameters are shown with the 
parametrization '$1b$', where $m_{f_0^L}=402$~MeV. In this case $T_c$ is
between $150-200$~MeV, which is in the range of the lattice results.

\begin{figure}[t!]
  \centering
  \begin{minipage}{.48\textwidth}
    \centering
    \includegraphics[width=1.05\textwidth]{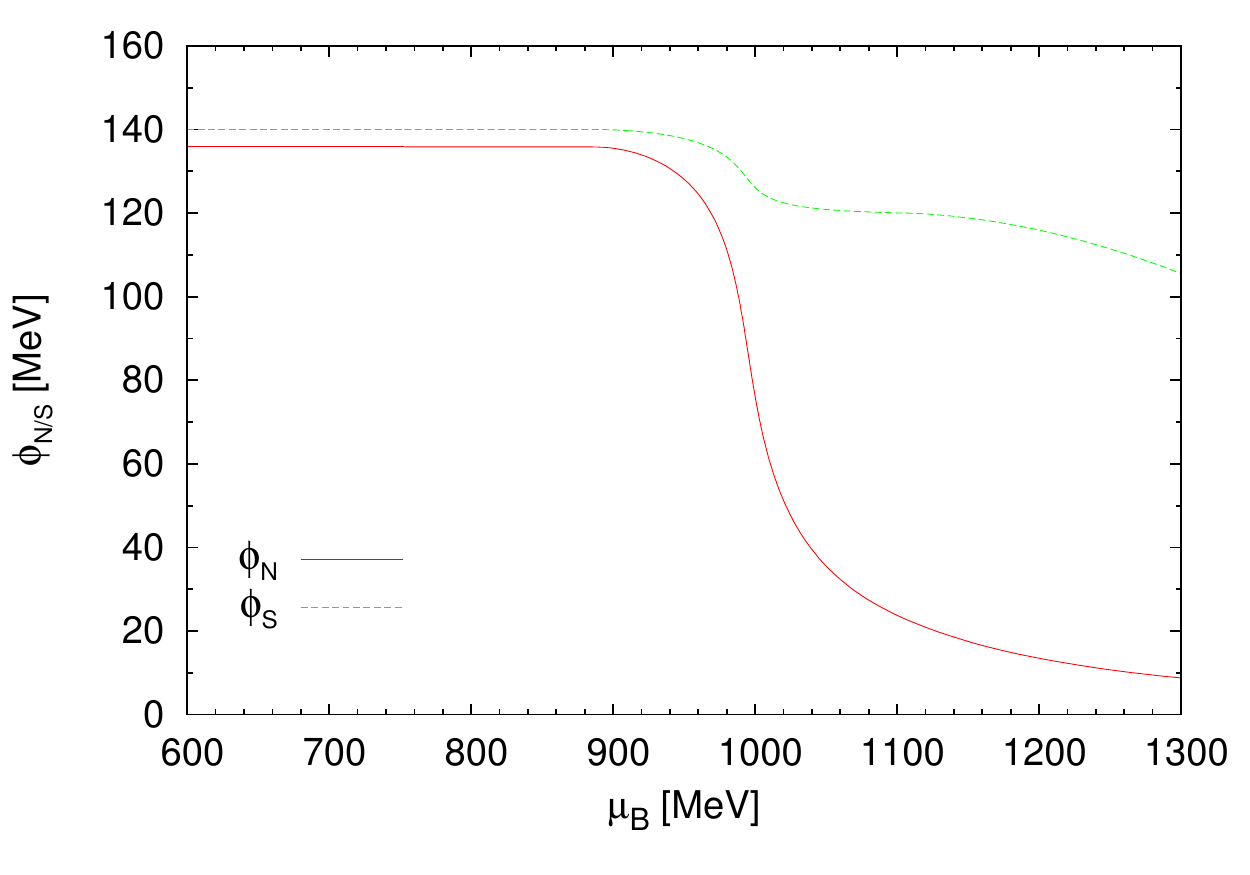}
    \caption{$\mu_B$ dependence of the condensates with the 
      parametrization '$1b$'.}
    \label{order_muB_old}
  \end{minipage}
  \hspace*{0.02\textwidth}
  \begin{minipage}{.48\textwidth}
    \centering
    \includegraphics[width=1.05\textwidth]{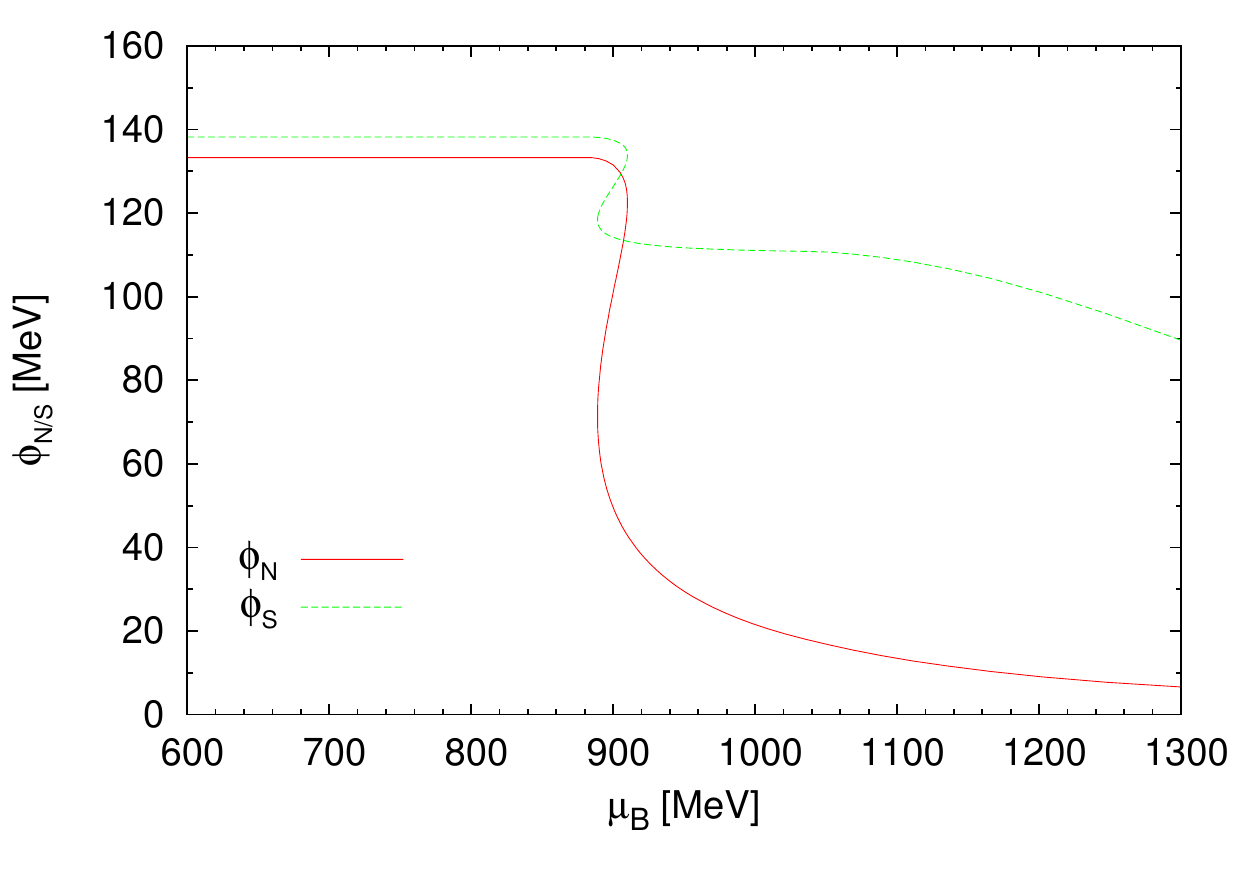}
    \caption{$\mu_B$ dependence of the condensates with the 
      parametrization '$2$'.}
    \label{order_muB_new}
  \end{minipage}
\end{figure}

\begin{figure}[t!]
  \centering
  \begin{minipage}{.48\textwidth}
    \centering
    \includegraphics[width=1.05\textwidth]{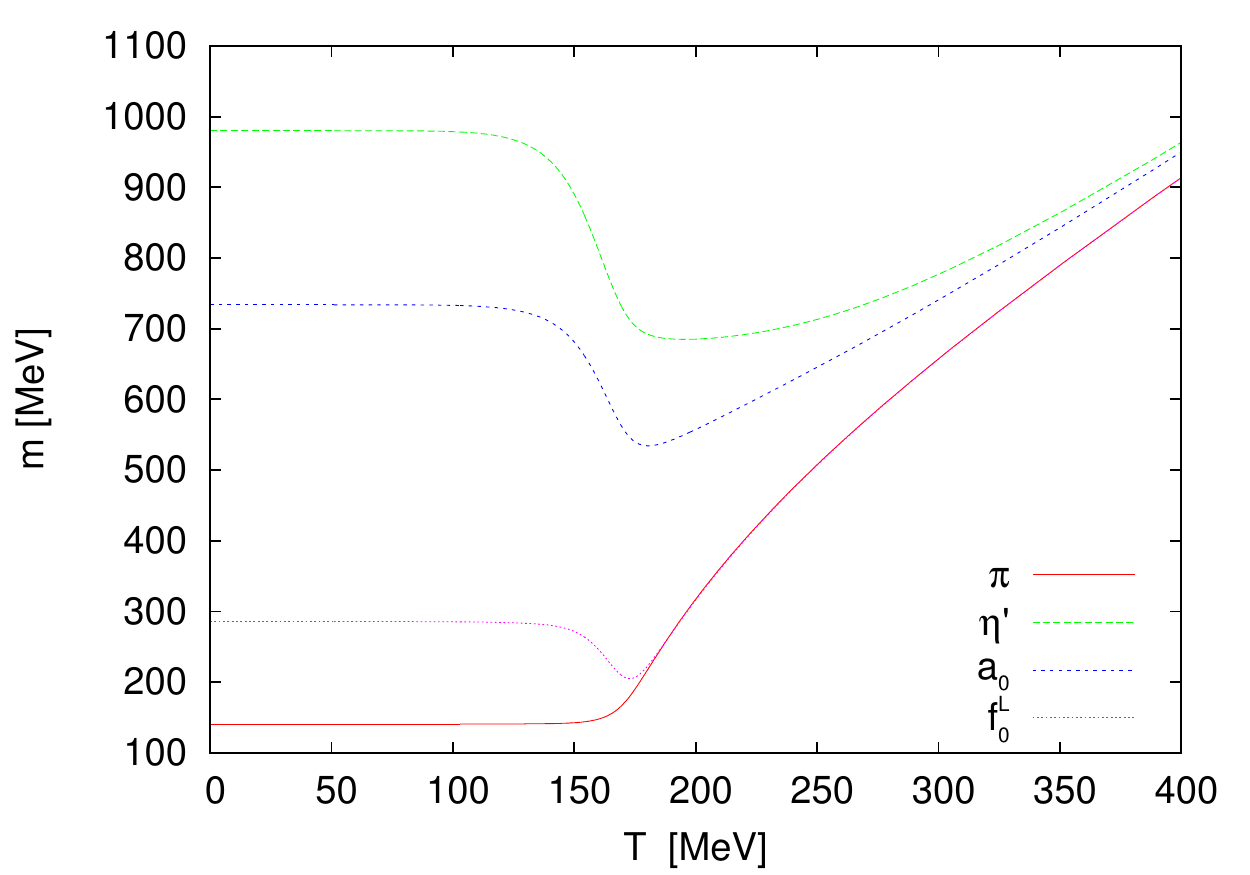}
    \caption{Temperature dependence of the $\pi$, $\eta^{\prime}$,
      $a_0$ and $f_0^L( i.e.\  \sigma)$ particle masses.}
    \label{mass_pi}
  \end{minipage}
  \hspace*{0.02\textwidth}
  \begin{minipage}{.48\textwidth}
    \centering
    \includegraphics[width=1.05\textwidth]{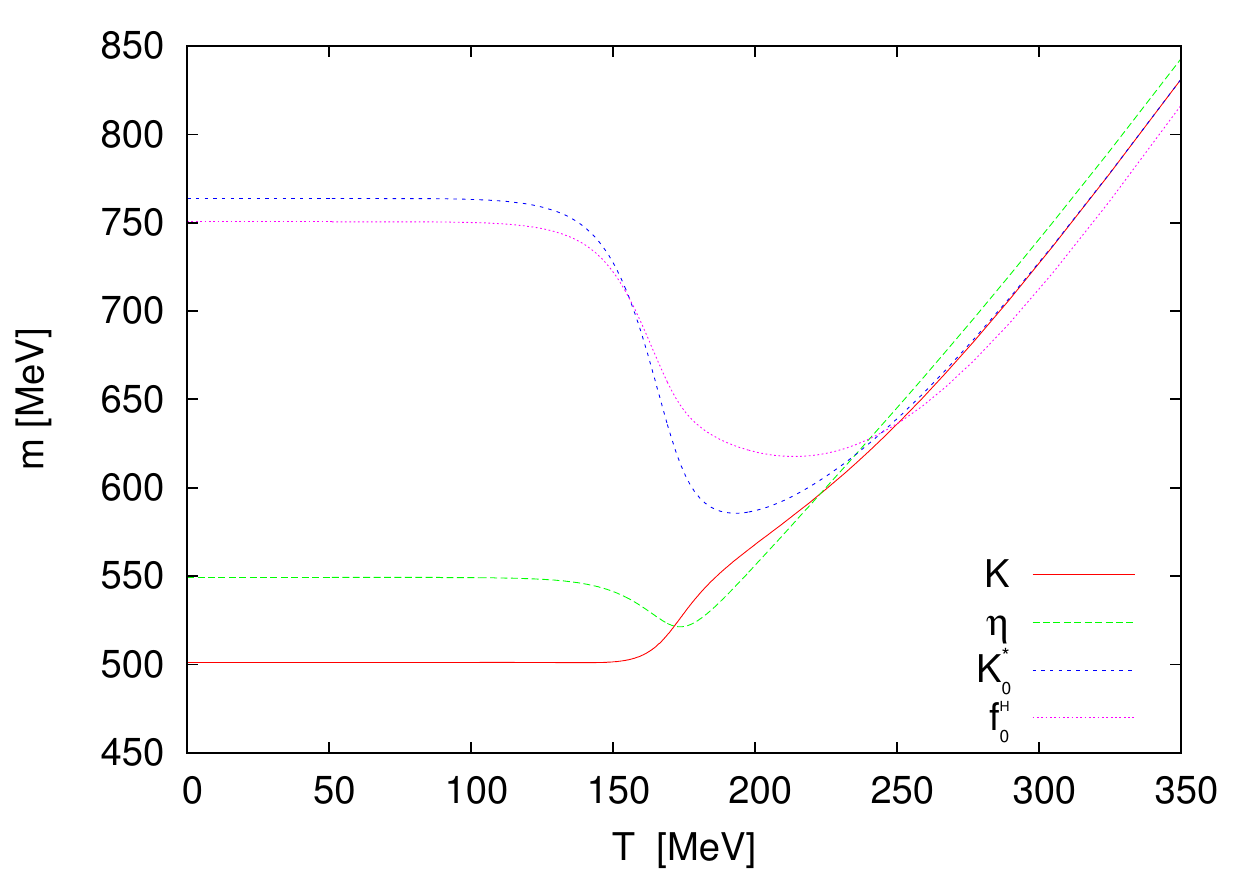}
    \caption{Temperature dependence of the $K$, $\eta$, $K_0^{\star}$
      and $f_0^{H}$ particle masses.}
    \label{mass_K}
  \end{minipage}
\end{figure}

It is a common belief that there is a critical endpoint (CEP) along
the phase boundary in the $T-\mu_B$ plane. For this to happen, the
phase transition should be of first order as a function of $\mu_B$
along the $T=0$ axis.  However, as can be seen in
Fig.~\ref{order_muB_old} (case '$1b$') the phase transition at $T=0$
is of crossover type, while in Fig.~\ref{order_muB_new} with
parametrization '$2$' the transition is of first order.  In this case
the value of the pseudocritical temperature $T_c$ at $\mu_B=0$ is
close to that of case '$1b$'.  The temperature dependence of the
pseudo(scalar) curvature masses can be seen in Fig.~\ref{mass_pi} and
Fig.~\ref{mass_K} for the parametrization of case '$2$'.

In conclusion we can say that the parametrization which fulfills
the two physical requirements of having a pseudocritical
temperature $T_c\approx 150$~MeV at $\mu_B=0$ and a first order
transition in $\mu_B$ at $T=0$ is realized in case '$2$', that is when
$f_0(500)$ and $f_0(1370)$ are used for parametrization.
Similar results were found in
\cite{tiwari_2013, chatterjee_2012} without (axial)vector mesons. 


% \section{Conclusion}
% \label{Sec:conclusion}

\ack

The authors were supported by the Hungarian OTKA fund K109462 and by the
HIC for FAIR Guest Funds of the Goethe University Frankfurt.

\end{document}